\makeatletter
\declare@file@substitution{revtex4-1.cls}{revtex4-2.cls}
\makeatother

\documentclass{aastex62}
\pdfoutput=1
\submitjournal{ApJ}

\begin{document}

\title{Andromeda's Parachute: Time Delays and Hubble Constant}
\shorttitle{Andromeda's Parachute: Time Delays and Hubble Constant}

\correspondingauthor{Luis J. Goicoechea}
\email{goicol@unican.es}

\author{Vyacheslav N. Shalyapin}
\affil{Departamento de F\'\i sica Moderna, Universidad de Cantabria, Avda. de Los Castros s/n, 39005 Santander, Spain}
\affil{O.Ya. Usikov Institute for Radiophysics and Electronics, NASU, 12 Acad. Proscury St., 61085 Kharkiv, Ukraine}
\affil{Institute of Astronomy of V.N. Karazin Kharkiv National University, Svobody Sq. 4, 61022 Kharkiv, Ukraine}

\author{Luis J. Goicoechea}
\affil{Departamento de F\'\i sica Moderna, Universidad de Cantabria, Avda. de Los Castros s/n, 39005 Santander, Spain}

\author{Karianne Dyrland}
\affil{Institute of Theoretical Astrophysics, University of Oslo, PO Box 1029, Blindern 0315, Oslo, Norway}
\affil{Kongsberg Defence \& Aerospace AS, Instituttveien 10, PO Box 26, 2027, Kjeller, Norway}

\author{H\r{a}kon Dahle}
\affil{Institute of Theoretical Astrophysics, University of Oslo, PO Box 1029, Blindern 0315, Oslo, Norway}



\begin{abstract}
The gravitational lens system PS J0147+4630 (Andromeda's Parachute) consists 
of four quasar images ABCD and a lensing galaxy. We obtained $r$--band light curves of ABCD 
in the 2017$-$2022 period from monitoring with two 2--m class telescopes. Applying 
state--of--the--art curve shifting algorithms to these light curves led to measurements of 
time delays between images, and the three independent delays relative to image D are 
accurate enough to be used in cosmological studies (uncertainty of about 4\%): $\Delta 
t_{\rm{AD}}$ = $-$170.5 $\pm$ 7.0, $\Delta t_{\rm{BD}}$ = $-$170.4 $\pm$ 6.0, and $\Delta 
t_{\rm{CD}}$ = $-$177.0 $\pm$ 6.5 d, where image D is trailing all the other images. Our 
finely sampled light curves and some additional fluxes in the years 2010$-$2013 also 
demonstrated the presence of significant microlensing variations. From the measured delays 
relative to image D and typical values of the external convergence, recent lens mass models
yielded a Hubble constant that is in clear disagreement with currently accepted values 
around 70 km s$^{-1}$ Mpc$^{-1}$. We discuss how to account for a standard value of the 
Hubble constant without invoking the presence of an extraordinary high external convergence.   
\end{abstract}



\section{Introduction} \label{sec:intro}

Optical frames from the Panoramic Survey Telescope and Rapid Response System 
\citep[Pan--STARRS;][]{2016arXiv161205560C} led to the serendipitous discovery of the strong 
gravitational lens system with a quadruply--imaged quasar (quad) \object{PS J0147+4630} 
\citep{2017ApJ...844...90B}. Due to its position in the sky and the spatial arrangement of 
the four quasar images, this quad is also called Andromeda's Parachute 
\citep[e.g.,][]{2018ApJ...859..146R}. The three brightest images (A, B and C) form an arc 
that is about 3\arcsec\ from the faintest image D, and the main lens galaxy G is located 
between the bright arc and D. This configuration is clearly seen in the left panel of 
Figure~\ref{fig:f1}, which is based on {\it Hubble Space Telescope} ($HST$) data. 

Early optical spectra of the system confirmed the gravitational lensing phenomenon and 
revealed the broad absorption--line nature of the quasar, which has a redshift $z_{\rm{s}} 
\sim$ 2.36 \citep{2017A&A...605L...8L,2018ApJ...859..146R}. \citet{2018MNRAS.475.3086L} also 
performed the first attempt to determine the redshift of G from spectroscopic observations 
with the 8.1 m Gemini North Telescope (GNT). An accurate reanalysis of these GNT data showed 
that the first estimate of the lens redshift was biased, by enabling better identification 
of G as an early--type galaxy at $z_{\rm{l}}$ = 0.678 $\pm$ 0.001 with stellar velocity 
dispersion $\sigma_{\rm{l}}$ = 313 $\pm$ 14 km s$^{-1}$ \citep{2019ApJ...887..126G}, in good 
agreement with the recent measurements of $z_{\rm{l}}$ and $\sigma_{\rm{l}}$ by 
\citet{2023A&A...672A..20M}. 

As far as we know, the quasar \object{PS J0147+4630} is the brightest source in the sky at 
redshifts $z >$ 1.4 (apart from transient events such as gamma--ray bursts), and its four 
optical images can be easily resolved with a ground--based telescope in normal seeing 
conditions. Thus, it is a compelling target for various physical studies based on 
high--resolution spectroscopy \citep[e.g.,][]{2018ApJ...859..146R} and detailed photometric 
monitoring \citep[e.g.,][]{2018MNRAS.475.3086L}. Early two--season monitoring campaigns with 
the 2.0 m Liverpool Telescope \citep[LT;][]{2019ApJ...887..126G} and the 2.5 m Nordic 
Optical Telescope \citep[NOT;][]{MScKD} provided accurate optical light curves of all quasar 
images, as well as preliminary time delays and evidence of microlensing--induced variations. 
A deeper look at the optical variability of Andromeda's Parachute is of great importance, 
since robust time delays and well--observed microlensing variations can be used to determine 
cosmological parameters \citep[e.g.,][]{2016A&ARv..24...11T} and the structure of the quasar 
accretion disc \citep[e.g.,][]{2010GReGr..42.2127S}. 

This paper is organized as follows. In Sect.~\ref{sec:lcs}, we present combined LT and NOT 
light curves of the four images of \object{PS J0147+4630} spanning six observing seasons 
from 2017 to 2022. In Sect.~\ref{sec:delmic}, using these optical light curves, we carefully 
analyse the time delays between images and the quasar microlensing variability. In 
Sect.~\ref{sec:mass}, we discuss the Hubble constant ($H_0$) value from the measured time 
delays and lens mass models. Our main conclusions are included in Sect.~\ref{sec:end}.

\section{New optical light curves} \label{sec:lcs} 

We monitored \object{PS J0147+4630} with the LT from 2017 August to 2022 October using the 
IO:O optical camera with a pixel scale of $\sim$0\farcs30. Each observing night, a single 
120 s exposure was taken in the Sloan $r$--band filter, and over the full monitoring period, 
212 $r$--band frames were obtained. The LT data reduction pipeline carried out three basic 
tasks: bias subtraction, overscan trimming, and flat fielding. Additionally, the IRAF 
software\footnote{https://iraf-community.github.io/} \citep{1986SPIE..627..733T,
1993ASPC...52..173T} allowed us to remove cosmic rays and bad pixels from all frames. We 
extracted the brightness of the four quasar images ABCD through PSF fitting, using the 
IMFITFITS software \citep{1998AJ....115.1377M} and following the scheme described by 
\citet{2019ApJ...887..126G}. Table~\ref{tab:t1} includes the position and magnitudes of the 
PSF star, as well as of other relevant field stars. These data are taken from the Data 
Release 1 of Pan--STARRS\footnote{http://panstarrs.stsci.edu} \citep{2020ApJS..251....7F}. 
Our photometric model consisted of four point--like sources (ABCD) and a de Vaucouleurs 
profile convolved with the empirical PSF (lensing galaxy G). Positions of components with 
respect to A and structure parameters of G were constrained from $HST$ data 
\citep{2019MNRAS.483.5649S,2021MNRAS.501.2833S}.

\begin{deluxetable}{lccccc}[h!]
\tablecaption{Pan--STARRS positions and magnitudes of relevant field stars.\label{tab:t1}}
\tablenum{1}
\tablewidth{0pt}
\tablehead{
\colhead{Star} & 
\colhead{RA(J2000)} &  
\colhead{Dec(J2000)} & 
\colhead{$g$} &
\colhead{$r$} & 
\colhead{$i$}  
}
\startdata
PSF 	& 26.773246 & 46.506670 & 16.366 & 15.606 & 15.260 \\
S       & 26.746290 & 46.504028 & 15.800 & 15.421 & 15.269 \\
Cal1    & 26.805695 & 46.522834 & 16.587 & 16.292 & 16.208 \\
Cal2    & 26.725610 & 46.488113 & 16.857 & 16.405 & 16.257 \\
Cal3    & 26.752831 & 46.518659 & 17.157 & 16.836 & 16.718 \\
Cal4    & 26.760809 & 46.474513 & 17.229 & 16.856 & 16.714 \\
Cal5    & 26.824027 & 46.528718 & 15.656 & 15.200 & 15.029 \\
Cal6    & 26.790480 & 46.502241 & 15.145 & 14.831 & 14.716 \\
\enddata
\tablecomments{Astrometric and photometric data of the stars that we used for PSF fitting 
(PSF), variability checking (control star S), and calibration (Cal1--Cal6). RA(J2000) and 
Dec(J2000) are given in degrees.}
\end{deluxetable}

\begin{deluxetable}{cccccccccccc}[h!]
\tablecaption{New $r$--band light curves of PS J0147+4630ABCD and the control star S.
\label{tab:t2}}
\tablenum{2}
\tablewidth{0pt}
\tablehead{
\colhead{Epoch\tablenotemark{a}} & 
\colhead{A\tablenotemark{b}} & \colhead{err(A)\tablenotemark{b}} &
\colhead{B\tablenotemark{b}} & \colhead{err(B)\tablenotemark{b}} &
\colhead{C\tablenotemark{b}} & \colhead{err(C)\tablenotemark{b}} &
\colhead{D\tablenotemark{b}} & \colhead{err(D)\tablenotemark{b}} &
\colhead{S\tablenotemark{b}} & \colhead{err(S)\tablenotemark{b}} &
\colhead{Tel\tablenotemark{c}}
}
\startdata
7970.051 & 15.945 & 0.005 & 16.174 & 0.007 & 16.616 & 0.008 & 18.188 & 0.017 & 15.410 & 0.005 & LT \\
7976.081 & 15.944 & 0.008 & 16.189 & 0.009 & 16.613 & 0.012 & 18.201 & 0.024 & 15.412 & 0.007 & LT \\
7982.116 & 15.961 & 0.006 & 16.195 & 0.007 & 16.628 & 0.009 & 18.228 & 0.018 & 15.413 & 0.005 & LT \\
7985.157 & 15.948 & 0.012 & 16.191 & 0.012 & 16.608 & 0.014 & 18.221 & 0.019 & 15.396 & 0.017 & NOT\\
7991.048 & 15.956 & 0.006 & 16.204 & 0.007 & 16.630 & 0.009 & 18.234 & 0.018 & 15.410 & 0.005 & LT \\
\enddata
\tablenotetext{a}{MJD--50000.}
\tablenotetext{b}{$r$--SDSS magnitude.}
\tablenotetext{c}{Tel indicates the used telescope (LT or NOT).}
\tablecomments{Table 2 is published in its entirety in the machine--readable format.
A portion is shown here for guidance regarding its form and content.}
\end{deluxetable}

\begin{figure}[h!]
	\begin{minipage}[h]{0.3\linewidth}
      \centering
      \includegraphics[width=1.0\textwidth]{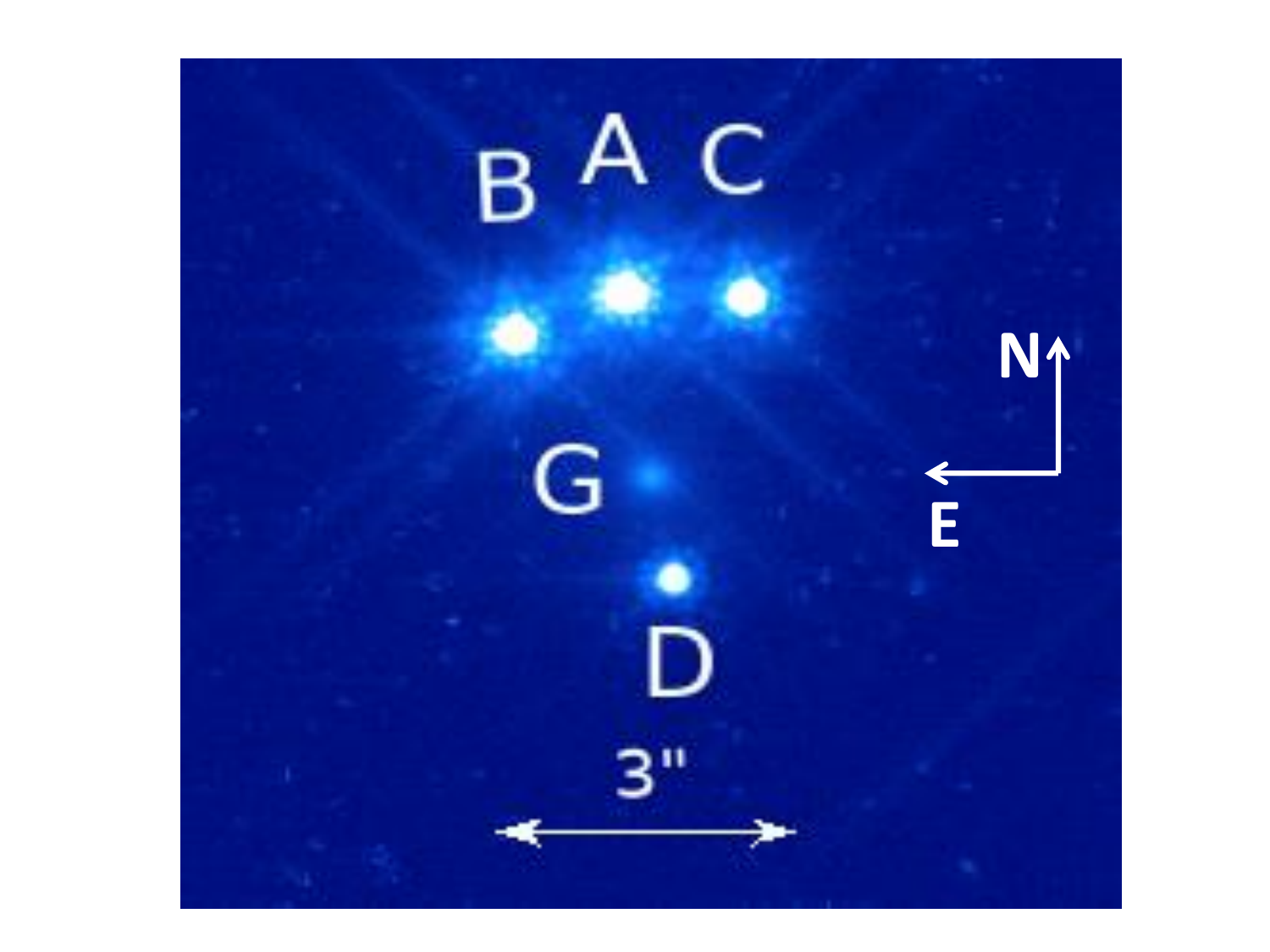}
        \end{minipage}
	\begin{minipage}[h]{0.7\linewidth}
      \centering
      \includegraphics[width=1.0\textwidth]{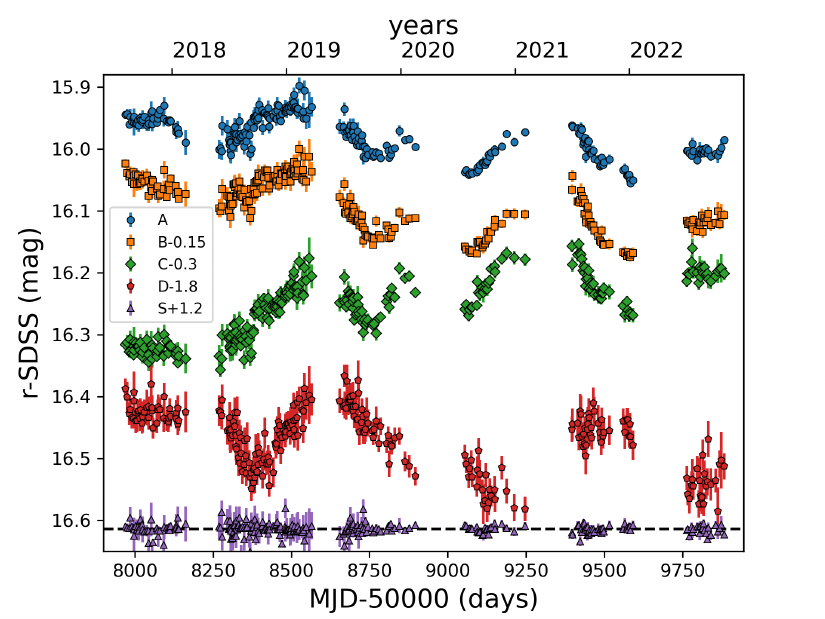}
    	\end{minipage}
\caption{Left: Quasar images ABCD and main lens galaxy G of PS J0147+4630 from a public 
$HST$--WFC3 frame of the system in the $F814W$ band. Right: LT--NOT light curves of PS 
J0147+4630 from its discovery to 2022 October 30. The $r$--band magnitudes of images B, 
C, and D, and the control star are offset by $-$0.15, $-$0.3, $-$1.8, and +1.2, 
respectively, to facilitate comparison between them and with image A.}
\label{fig:f1}
\end{figure}

We also selected six non--variable blue stars in the field of \object{PS J0147+4630} and 
performed PSF photometry on five of them (see the calibration stars Cal1--Cal5 in 
Table~\ref{tab:t1}; Cal6 is a saturated star in LT frames). For each of the five calibration 
stars, we calculated its average magnitude within the monitoring period and magnitude 
deviations in individual frames (by subtracting the average). In each individual frame, the 
five stellar magnitude deviations were averaged together to calculate a single magnitude 
offset, which was then subtracted from the magnitudes of quasar images. After this 
photometric calibration, we removed 22 observing epochs in which quasar magnitudes deviate 
appreciably from adjacent values. Thus, the final LT $r$--band light curves are based on 190 
frames (epochs), and the typical uncertainties in the light curves of the quasar images and 
control star (see Table~\ref{tab:t1}) were estimated from magnitude differences between 
adjacent epochs separated by no more than 4.5 d \citep{2019ApJ...887..126G}. We derived 
typical errors of 0.0062 (A), 0.0077 (B), 0.0097 (C), 0.0197 (D), and 0.0058 (control star) 
mag. For the control star, we have also verified that its typical error practically 
coincides with the standard deviation of all measures (0.0055 mag). To obtain photometric 
uncertainties at each observing epoch, the typical errors were scaled by the relative 
signal--to--noise ratio of the PSF star \citep{2000hccd.book.....H}.    

The optical monitoring of \object{PS J0147+4630} with the NOT spanned from 2017 August to 
2019 December. We used the ALFOSC camera with a pixel scale of $\sim$0\farcs21 and the 
$R$--Bessel filter. This passband is slightly redder than the Sloan $r$ band. Each observing 
night, we mainly took three exposures of 30 s each under good seeing conditions. The 
full--width at half--maximum (FWHM) of the seeing disc was about 1\farcs0 (we also estimated 
FWHM seeing = 1\farcs35 $\pm$ 0\farcs15 from LT frames), and we collected 298 individual 
frames over the entire monitoring campaign. After a standard data reduction, IMFITFITS PSF 
photometry yielded magnitudes for the quasar images (see above for details on the 
photometric model). To avoid biases in the combined LT--NOT light curves, the same 
photometric method was applied to LT and NOT frames. This method differs from that of 
\citet{MScKD}, who used the DAOPHOT package in IRAF \citep{1987PASP...99..191S,TRMD} to 
extract magnitudes from NOT frames. 

The six calibration stars in Table~\ref{tab:t1} were used to adequately correct quasar 
magnitudes (see above), and we were forced to remove 17 individual frames leading to 
magnitude outliers. We then combined $R$--band magnitudes measured on the same night to 
obtain photometric data of the lensed quasar and control star at 77 epochs. Again, typical 
errors were derived from magnitudes at adjacent epochs that are separated $<$ 4.5 d. This 
procedure led to uncertainties of 0.0122 (A), 0.0122 (B), 0.0144 (C), 0.0197 (D), and 0.0170 
(control star) mag. Errors at each observing epoch were calculated in the same way as for 
the LT light curves. 

As a last step, we combined the $r$--band LT and $R$--band NOT light curves. If we focus on 
the quasar images and consider $rR$ pairs separated by no more than 2.5 d, the values of the 
average colour $\langle r - R \rangle$ are 0.0565 (A), 0.0616 (B), 0.0546 (C), and 0.0652 
(D). Brightness records of the ABC images are more accurate than those of D, and thus we 
reasonably take the average colours of ABC to estimate a mean $r - R$ offset of 0.0576 mag. 
After correcting the $R$--band curves of the quasar for this offset, we obtain the new 
records in Table~\ref{tab:t2}. Table~\ref{tab:t2} contains $r$--band magnitudes of the 
quasar images and the control star at 267 observing epochs (MJD$-$50\,000). In 
Figure~\ref{fig:f1}, we also display our new 5.2--year light curves.

\section{Time delays and microlensing signals} \label{sec:delmic}

Previous efforts focused on early monitorings with a single telescope, trying to estimate 
delays between the image A and the other quasar images, $\Delta t_{\rm{AX}} = t_{\rm{A}} - 
t_{\rm{X}}$ (X = B, C, D), and find microlensing signals 
\citep{MScKD,2019ApJ...887..126G}\footnote{\citet{2019ApJ...887..126G} used the notation 
$\Delta t_{\rm{AX}} = t_{\rm{X}} - t_{\rm{A}}$ rather than that defined in this paper and 
\citet{MScKD}}. Here, we use the new light curves in Section \ref{sec:lcs} along with 
state--of--the--art curve--shifting algorithms to try to robustly measure time delays 
between images. At the end of this section, we also discuss the extrinsic (microlensing) 
variability of the quasar.  

As is clear from Figure~\ref{fig:f1}, there are short time delays between images ABC, while 
it is hard to get an idea about the $\Delta t_{\rm{AD}}$ value by eye. Fortunately, there 
are several cross--correlation techniques to measure time delays between light curves 
containing microlensing variations \citep[e.g.,][and references 
therein]{2015ApJ...800...11L}, and thus we considered PyCS3 curve--shifting 
algorithms\footnote{https://gitlab.com/cosmograil/PyCS3} \citep{2013A&A...553A.120T,PyCS3,
2020A&A...640A.105M} to obtain reliable time delays of \object{PS J0147+4630}. PyCS3 is a
well--tested software toolbox to estimate time delays between images of gravitationally 
lensed quasars, and we focused on the $\chi^2$ technique, assuming that the intrinsic signal 
and the extrinsic ones can be modelled as a free--knot spline (FKS). This technique shifts 
the four light curves simultaneously (ABCD comparison) to better constrain the intrinsic 
variability, and relies on an iterative nonlinear procedure to fit the four time shifts and 
splines that minimise the $\chi^2$ between the data and model \citep{2013A&A...553A.120T}. 
Results depend on the initial guesses for the time shifts, so it is necessary to estimate 
the intrinsic variance of the method using a few hundred initial shifts randomly distributed 
within reasonable time intervals. In addition, a FKS is characterised by a knot step, which 
represents the initial spacing between knots. The model consists of an intrinsic spline with 
a knot step $\eta$ and four independent extrinsic splines with $\eta_{\rm{ml}}$ that account 
for the microlensing variations in each quasar image \citep{2020A&A...640A.105M}. 

\begin{figure}[h!]
\centering
\includegraphics[width=\textwidth]{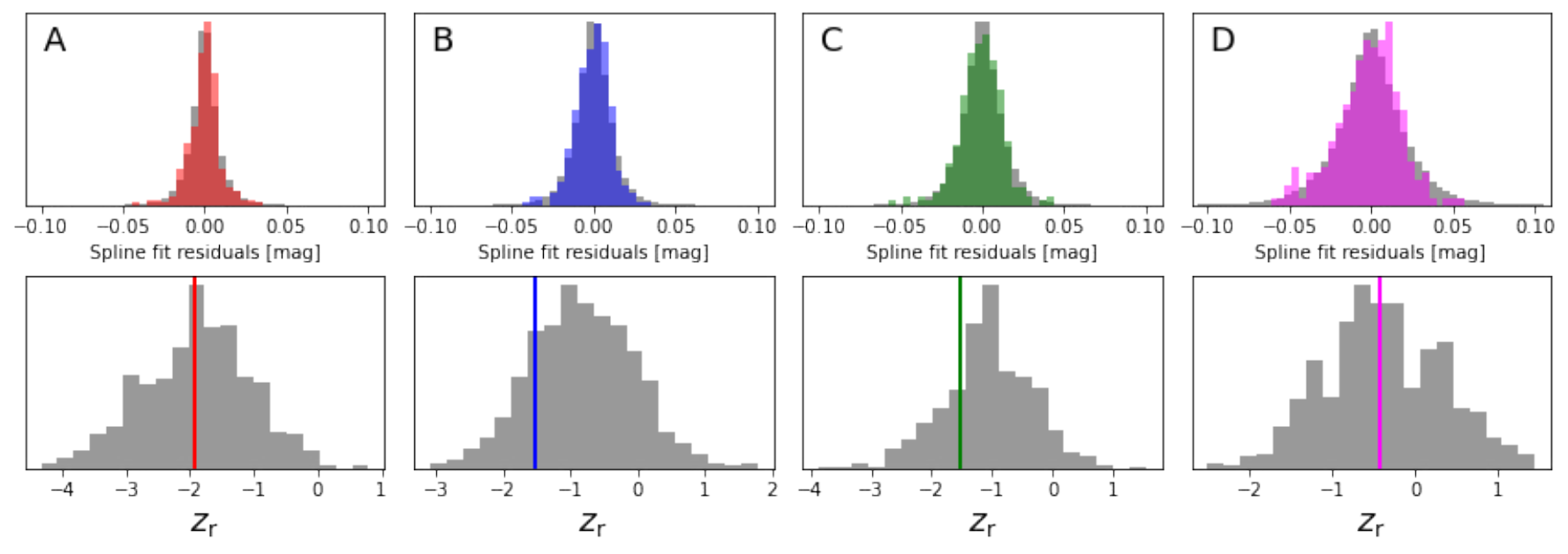}
\caption{Top: Distributions of FKS fit residuals for $\eta$ = 50 d and $\eta_{\rm{ml}}$ = 
400 d. The grey histograms represent the distributions of residuals from the fits to 500 
synthetic light curves of each image, while the red, blue, green and magenta histograms 
correspond to the distributions of residuals from the fits to the LT--NOT light curves. 
Bottom: Normalised number of runs $Z_{\rm{r}}$ for the synthetic data (grey histograms) and 
the observed brightness records (red, blue, green and magenta vertical lines).}
\label{fig:f2}
\end{figure}

To address the intrinsic variability, we considered three $\eta$ values of 30, 50 and 70 d. 
Intrinsic knot steps shorter than 30 d fit the observational noise, whereas $\eta$ values 
longer than 70 d do not fit the most rapid variations of the source quasar. Intrinsic 
variations are usually faster than extrinsic ones, and additionally, the software works fine 
when the microlensing knot step is significantly longer than $\eta$. Therefore, the 
microlensing signals were modelled as free--knot splines with $\eta_{\rm{ml}}$ = 350$-$400 d 
\citep[i.e., values intermediate between those shown in Table 2 of][]{2020A&A...640A.105M}. 
We also generated 500 synthetic (mock) light curves of each quasar image, optimised every 
mock ABCD dataset, and checked the similarity between residuals from the fits to the 
observed curves and residuals from the fits to mock curves. The comparison of residuals was 
made by means of two statistics: standard deviation and normalised number of runs 
$Z_{\rm{r}}$ \citep[see details in][]{2013A&A...553A.120T}. For $\eta$ = 50 d and 
$\eta_{\rm{ml}}$ = 400 d, histograms of residuals derived from mock curves (grey) and from 
the LT--NOT light curves of \object{PS J0147+4630} are included in the top panels of 
Figure~\ref{fig:f2}. It is apparent that the standard deviations through the synthetic and 
the observed curves match very well. Additionally, the bottom panels of Figure~\ref{fig:f2} 
show distributions of $Z_{\rm{r}}$ from synthetic light curves (grey) for $\eta$ = 50 d and 
$\eta_{\rm{ml}}$ = 400 d. These bottom panels also display the $Z_{\rm{r}}$ values from the 
observations (vertical lines), which are typically located at $\sim$0.4$\sigma$ of the mean 
values of the synthetic distributions.

\begin{figure}[h!]
\centering
\includegraphics[width=0.84\textwidth]{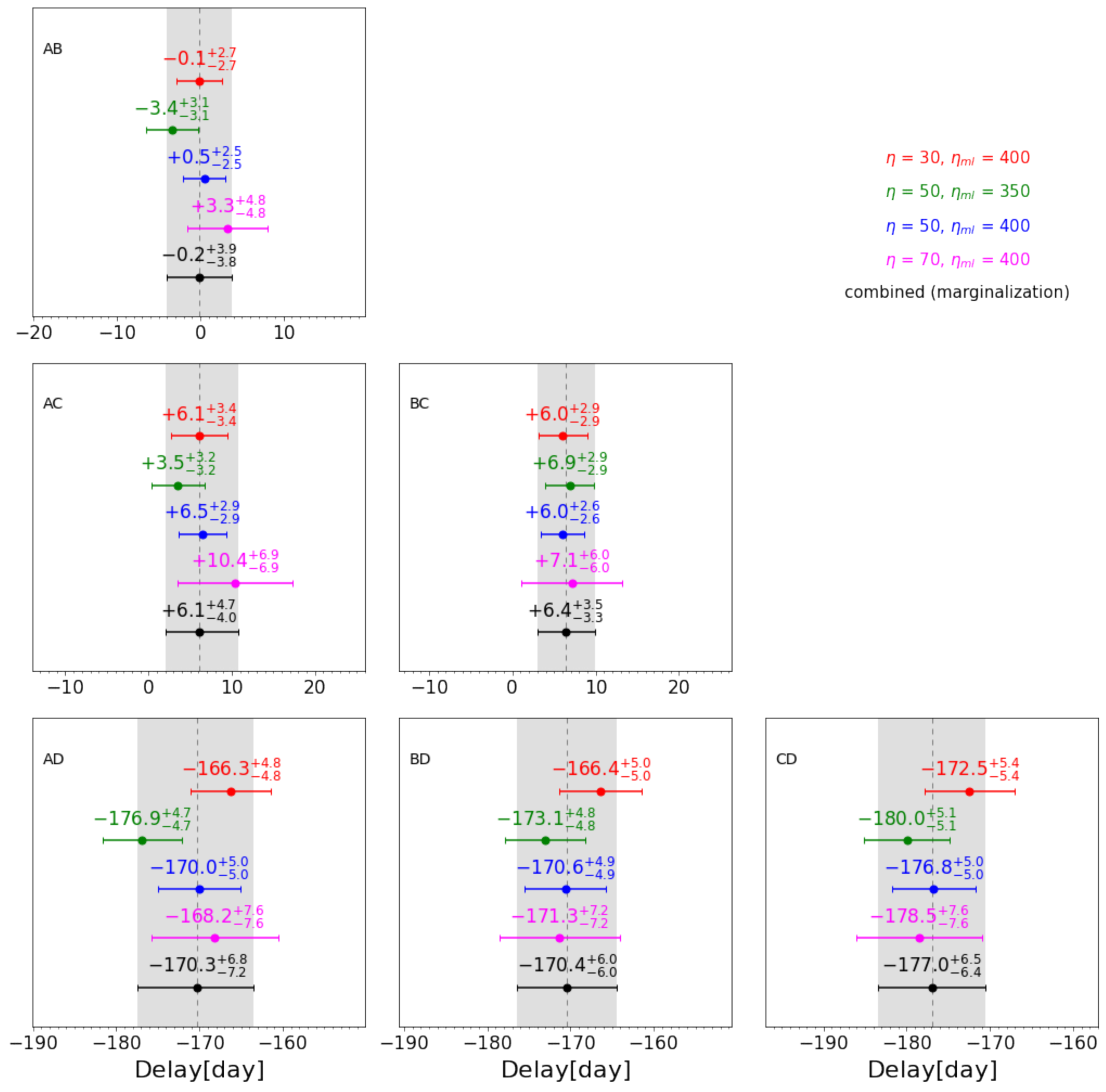}
\caption{Time--delay estimates using the $\chi^2$ technique and free--knot splines. Combined 
estimates \citep[$\tau_{\rm{thresh}}$ = 0;][]{2018A&A...616A.183B} are highlighted with grey 
rectangles encompassing the individual measurements.}
\label{fig:f3}
\end{figure}

Four pairs of ($\eta$, $\eta_{\rm{ml}}$) values (see above) led to the set of time delays in 
Figure~\ref{fig:f3}. We have verified that other feasible choices for $\eta_{\rm{ml}}$ (e.g., 
$\eta_{\rm{ml}}$ = 200 d) do not substantially modify the results in this figure. The black 
horizontal bars correspond to 1$\sigma$ confidence intervals after a marginalisation over 
results for all pairs of knot steps, and those in the left panels and bottom panels of 
Figure~\ref{fig:f3} are included in Table~\ref{tab:t3}. We finally adopted the time delays 
in Table~\ref{tab:t3}, which are symmetric about central values and useful for subsequent 
studies. 

It seems to be difficult to accurately determine delays between the brightest images ABC 
because they are really short. To robustly measure $\Delta t_{\rm{AC}}$ in a near future, we 
will most likely need to follow a non--standard strategy focused on several time segments 
associated with strong intrinsic variations and weak extrinsic signals. Fortunately, we find 
an accurate and reliable value of $\Delta t_{\rm{AD}}$ (uncertainty of about 4\%), 
confirming the early result from two monitoring seasons with the NOT and a technique 
different to that we used in this paper \citep{MScKD}. It is also worth mentioning that the 
dispersion method ignoring microlensing variations \citep[the simplest approximation with 
fewer free parameters;][]{1996A&A...305...97P} produces an optimal AD delay separated by 
only 10 days from that obtained with PyCS3. We also note that $\Delta t_{\rm{BD}}$ and 
$\Delta t_{\rm{CD}}$ have errors of 3.5$-$3.7\%, and thus we present accurate values of the 
three independent time delays relative to image D. 

\begin{deluxetable}{ccccc}[h!]
\tablecaption{Time delays of PS J0147+4630.\label{tab:t3}}
\tablenum{3}
\tablewidth{0pt}
\tablehead{
\colhead{$\Delta t_{\rm{AB}}$} & 
\colhead{$\Delta t_{\rm{AC}}$} &  
\colhead{$\Delta t_{\rm{AD}}$} &
\colhead{$\Delta t_{\rm{BD}}$} &  
\colhead{$\Delta t_{\rm{CD}}$}    
}
\startdata
$-$0.2 $\pm$ 3.9 & +6.4 $\pm$ 4.4 & $-$170.5 $\pm$ 7.0 & $-$170.4 $\pm$ 6.0 & $-$177.0 $\pm$ 6.5\\
\enddata
\tablecomments{Here, delays $\Delta t_{\rm{YX}} = t_{\rm{Y}} - t_{\rm{X}}$ are in days, 
image Y leads image X if $\Delta t_{\rm{YX}} <$ 0 (otherwise Y trails X), and all 
measurements are 68\% confidence intervals. We combined the individual measures of $\Delta 
t_{\rm{YX}}$ from PyCS3 (see Figure~\ref{fig:f3}) and then made symmetric error bars.}
\end{deluxetable}  

\begin{figure}[h!]
\centering
\includegraphics[width=0.7\textwidth]{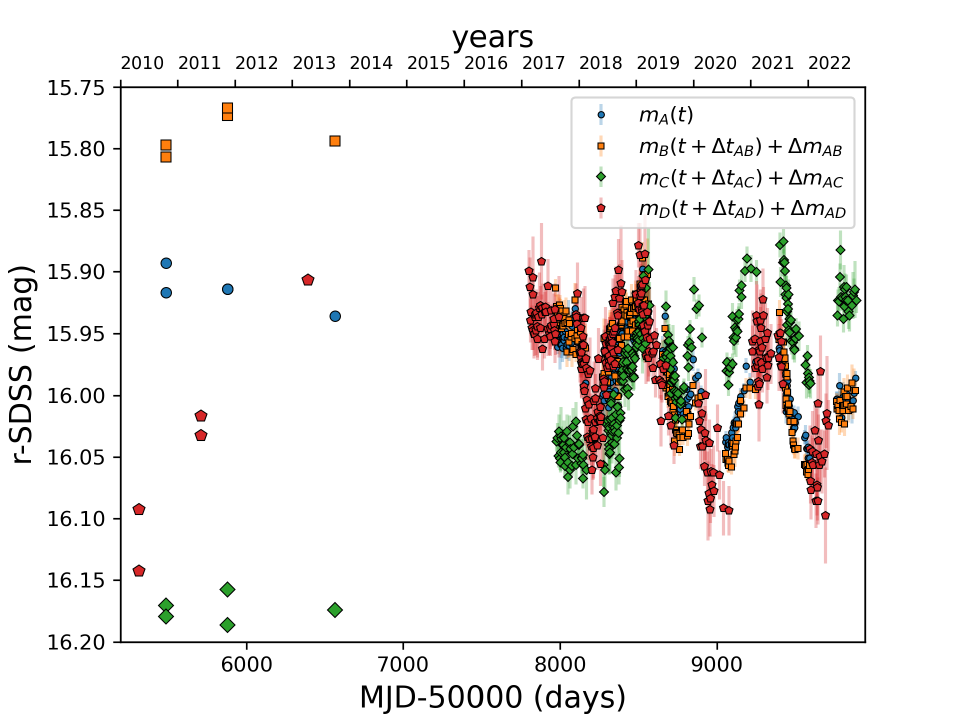}
\caption{LT--NOT data (smaller symbols) plus photometric data from Pan-STARRS $r$--band 
frames in 2010$-$2013 (larger symbols). The original brightness record of A is compared with 
shifted light curves of B, C, and D. To shift the BCD light curves, we apply the 
corresponding time delays and constant magnitude offsets (see main text for details).}
\label{fig:f4}
\end{figure}

An image comparison spanning 13 years is also depicted in Figure~\ref{fig:f4}. We have 
downloaded five $r$--band warp frames of \object{PS J0147+4630} that are included in the 
Data Release 2 of Pan--STARRS. These Pan--STARRS frames were obtained on three nights in the 
2010$-$2013 period, i.e., a few years before the discovery of the lens system. Two frames 
are available on two of the three nights, so rough photometric uncertainties through average 
intranight variations are 0.012 (A), 0.008 (B), 0.019 (C), and 0.033 (D) mag. To discuss the 
differential microlensing variability of the images BCD with respect to A, 
Figure~\ref{fig:f4} shows the original curve of A along with shifted curves of BCD. We used 
the central values of the delays relative to image A and constant magnitude offsets to shift 
curves. The offsets $\Delta m_{\rm{AB}}$, $\Delta m_{\rm{AC}}$, and $\Delta m_{\rm{AD}}$ are 
average differences between magnitudes of A and those of B, C, and D, respectively. The 
global shapes of the four brightness records indicate the presence of long--term 
microlensing effects and suggest that \object{PS J0147+4630} is a suitable target for a 
deeper analysis of its microlensing signals. Over the last six years, it is noteworthy that 
there is good overlap between the original curve of A and the shifted curve of D. In 
addition, the differential microlensing variation of C is particularly prominent, showing a 
microlensing episode with a total amplitude greater than 0.1 mag. 

\section{Lens mass models and Hubble constant} \label{sec:mass}

\citet{2017ApJ...844...90B} presented preliminary modelling of the lens mass of \object{PS 
J0147+4630} from Pan--STARRS data, whereas \citet{2019MNRAS.483.5649S,2021MNRAS.501.2833S} 
have recently modelled the lens system using $HST$ imaging. To model the $HST$ images, 
Shajib et al. have considered the distributions of light of the lens and source, and the 
lens mass distribution. Their solution for the lensing mass relies on a lens scenario 
consisting of a singular power--law ellipsoid (SPLE; describing the gravitational field of 
the main lens galaxy G) and an external shear (ES; accounting for the gravitational action 
of other galaxies). The dimensionless surface mass density (convergence) profile of the SPLE 
was characterised by a power--law index $\beta$ = 2.00 $\pm$ 0.05, where $\beta$ = 2 for an 
isothermal ellipsoid\footnote{The original notation for the power--law index in 
\citet{2019MNRAS.483.5649S,2021MNRAS.501.2833S} was $\gamma$, but we have renamed it as 
$\beta$ to avoid confusion between this index and the shear}. 

We first considered Shajib et al.'s solution, a flat $\Lambda$CDM (standard) cosmology with 
matter and dark energy densities of $\Omega_{\rm{M}}$ = 0.3 and $\Omega_{\Lambda}$ = 0.7, 
respectively\footnote{Results do not change appreciably for values of $\Omega_{\rm{M}}$ and 
$\Omega_{\Lambda}$ slightly different from those adopted here}, updated redshifts 
$z_{\rm{l}}$ = 0.678 \citep{2019ApJ...887..126G} and $z_{\rm{s}}$ = 2.357 (based on emission 
lines that are observed at near–-IR wavelengths), and the time delay in the third column of 
Table~\ref{tab:t3} to calculate $H_0^{\rm{model}}$ and put it into perspective. We obtained 
$H_0^{\rm{model}}$ = 100 $\pm$ 10 km s$^{-1}$ Mpc$^{-1}$, which significantly exceeds a 
concordance value of $\sim$70 km s$^{-1}$ Mpc$^{-1}$ \citep[e.g.,][]{2015LRR....18....2J}. 
If additional mass along the line of sight is modelled as an external convergence 
$\kappa_{\rm{ext}}$, then $H_0^{\rm{true}} = H_0^{\rm{model}} (1 - \kappa_{\rm{ext}})$ 
\citep[e.g.,][]{2017MNRAS.467.4220R}. The factor $1 - \kappa_{\rm{ext}}$ should be $\sim$0.7 
($\kappa_{\rm{ext}} \sim$ 0.3) to decrease $H_0$ until accepted values. Therefore, the 
external convergence required to solve the $H_0$ crisis is an order of magnitude higher than 
typical values of $\kappa_{\rm{ext}}$ \citep[e.g.,][]{2017MNRAS.467.4220R,
2020A&A...643A.165B}.

The Hubble constant can be also inferred from another lens mass solution based on approaches 
similar to those of Shajib et al. Adopting a standard cosmology and updated redshifts (see 
above), the solution of \citet{2023MNRAS.518.1260S} and the three time delays relative to 
image D (last three columns in Table~\ref{tab:t3}) led to $H_0^{\rm{model}}$ values in the 
range 116 to 131 km s$^{-1}$ Mpc$^{-1}$. Thus, Schmidt et al.'s solution with power--law 
index $\beta$ = 2.08 $\pm$ 0.02 produces even higher $H_0^{\rm{model}}$ values than those 
from Shajib et al.'s solution. Although the $H_0$ crisis may be related to an inappropriate 
(SPLE + ES) lens scenario or a very high external convergence, we have sought for a new mass 
reconstruction using astrometric and time--delay constraints, a SPLE + ES scenario, updated 
redshifts, a standard cosmology, and $H_0^{\rm{model}}$ = 70 km s$^{-1}$ Mpc$^{-1}$. In 
presence of a typical (weak) external convergence, the $H_0^{\rm{true}}$ value would be 
consistent with accepted ones.

Our standard astrometric constraints consisted of the $HST$ positions of ABCD \citep[with 
respect to G at the origin of coordinates;][]{2019MNRAS.483.5649S,2021MNRAS.501.2833S}. SPLE 
+ ES mass models of quads usually indicate the existence of an offset between the centre of 
the SPLE and the light centroid of the galaxy \citep[e.g.,][]{2012A&A...538A..99S,
2019MNRAS.483.5649S,2021MNRAS.501.2833S}. Hence, instead of formal astrometric errors for G, 
we adopted $\sigma_x$ = $\sigma_y$ = 0\farcs04. This uncertainty level equals the 
root--mean--square of mass/light positional offsets for most quads in the sample of Shajib 
et al. In addition to astrometric data, the set of constraints incorporated the LT--NOT time 
delays relative to image D (see Table~\ref{tab:t3}). The number of observational constraints 
and the number of model parameters were 13 and 10, respectively. For three degrees of 
freedom, the GRAVLENS/LENSMODEL 
software\footnote{\url{http://www.physics.rutgers.edu/~keeton/gravlens/}} 
\citep{2001astro.ph..2340K,gravlensmanual} led to the 1$\sigma$ intervals in 
Table~\ref{tab:t4} ($\chi^2$ = 3.56 for the best fit). 

\begin{deluxetable}{cccccc}[h!]
\tablecaption{SPLE + ES mass model of PS J0147+4630.\label{tab:t4}}
\tablenum{4}
\tablewidth{0pt}
\tablehead{
\colhead{$\beta$} & 
\colhead{$b$ (\arcsec)} &  
\colhead{$e$} &
\colhead{$\theta_e$ (\degr)} &  
\colhead{$\gamma$} &    
\colhead{$\theta_{\gamma}$ (\degr)} 
}
\startdata
1.86 $\pm$ 0.07 & 1.878 $\pm$ 0.018 & 0.170 $\pm$ 0.045 & $-$70.8 $\pm$ 3.5 & 0.177 $\pm$ 0.019 & 10.8 $\pm$ 0.7 \\
\enddata
\tablecomments{We consider $HST$ astrometry and LT--NOT time delays as constraints. We also 
adopt updated redshifts, a standard cosmology, and $H_0^{\rm{model}}$ = 70 km s$^{-1}$ 
Mpc$^{-1}$. Position angles ($\theta_e$ and $\theta_{\gamma}$) are measured east of north, 
and $\beta$, $b$, $e$, and $\gamma$ denote power--law index, mass scale and ellipticity of 
the SPLE, and external shear strength, respectively. We show 68\% (1$\sigma$) confidence 
intervals.}
\end{deluxetable}

While our solution for the mass of the early--type galaxy G is characterised by a 
convergence a little shallower than isothermal ($\beta <$ 2; see Table~\ref{tab:t4}), Shajib 
et al.'s and Schmidt et al.'s solutions for the surface mass density are more centrally 
concentrated ($\beta \geq$ 2), suggesting this is a key reason to infer such high values of 
$H_0^{\rm{model}}$ from previous models \citep[e.g.,][]{1994RPPh...57..117R,
2004mmu..symp..117K,2015LRR....18....2J}. The only issue with all SPLE + ES mass models is 
the existence of a significant mass/light misalignment, i.e., the light and mass 
distributions of the lens galaxy do not match. This misalignment could be genuine or due to 
an oversimplification of the lens scenario \citep[e.g.,][]{2012A&A...538A..99S,
2016ApJ...820...43S,2021MNRAS.504.1340G}. Most early--type galaxies reside in overdense 
regions, so external tidal fields in their vicinity are expected to have relatively high 
amplitudes. External shear strengths for quads exceeding 0.1 are consistent with N--body 
simulations and semianalytic models of galaxy formation \citep{2003ApJ...589..688H}. Using a 
model consisting of a singular isothermal elliptical potential and external shear, 
\citet{2021ApJ...915....4L} have also shown that \object{PS J0147+4630} is a 
shear--dominated system.
     
\section{Conclusions} \label{sec:end}

In this paper, we performed a comprehensive analysis of the optical variability of the 
quadruply--imaged quasar \object{PS J0147+4630}. Well--sampled light curves from its 
discovery in 2017 to 2022 were used to robustly measure the three time delays relative to 
image D. However, these light curves did not allow us to accurately (in terms of fractional 
uncertainty) determine the very short time delays between the three bright images ABC 
forming a compact arc. Additionally, the microlensing--induced variation of the C image 
(with respect to A) was particularly large in the period 2017$-$2022. Combining our new 
brightness records with quasar fluxes from Pan--STARRS imaging in 2010$-$2013, the extended 
light curves also revealed significant long--term microlensing effects. A microlensing 
analysis of current data and future light curves from a planned optical multi--band 
monitoring is expected to lead to important constraints on the spatial structure of the 
quasar accretion disc \citep{2008A&A...490..933E,2008ApJ...673...34P,2020ApJ...905....7C,
2020A&A...637A..89G}.

From $HST$ imaging of the quad, \citet{2019MNRAS.483.5649S,2021MNRAS.501.2833S} and 
\citet{2023MNRAS.518.1260S} have carried out reconstruction of the lensing mass from an SPLE 
+ ES scenario. However, using updated redshifts of the source and lens (and assuming a 
standard cosmology), their mass reconstructions along with measured delays relative to image 
D led to an unacceptably large value of the Hubble constant. Although the integrated mass 
from objects along the line of sight to \object{PS J0147+4630} is still unknown, an 
unexpected (unusually high) external convergence is required to fix this $H_0$ issue. To try
to overcome the $H_0$ crisis, we have sought and found a new mass model that is consistent 
with astrometric and time--delay constraints, a typical external convergence, and currently 
accepted values for $H_0$ around 70 km s$^{-1}$ Mpc$^{-1}$ \citep[e.g., see Fig. 2 
of][]{2021CQGra..38o3001D}. Time delays are very sensitive to the slope of the mass profile 
of the main lens galaxy G \citep[e.g.,][]{2004mmu..symp..117K}, and the new model 
incorporates a surface mass density less centrally concentrated than previous ones.  

Alternatively, the SPLE + ES lens scenario might be an oversimplification of the actual one, 
since all SPLE + ES models indicate that there is a mass/light misalignment. While this 
misalignment may be true, it could also be due to the presence of non--modelled components 
such as substructures and/or companions of G \citep[e.g.,][]{2012A&A...538A..99S,
2021MNRAS.504.1340G}. Further refinement of the lens scenario along with an extension and 
improvement of the set of observational constraints (future deep photometry and spectroscopy 
is a pending task of special relevance) will contribute to an accurate determination of 
$H_0$ and other cosmological parameters \citep[e.g.,][]{2017MNRAS.465.4914B,
2020A&A...643A.165B}. The forthcoming Legacy Survey of Space and Time (LSST) at the Vera C. 
Rubin Observatory should provide the strong lens community with a strong increase in the 
number of known lensed quasars with measured time delays. To be able to utilise such a large 
increase in the statistical sample to provide correspondingly precise and accurate 
measurements of $H_0$, it is crucial to reliably identify the systems with more complex lens 
scenarios that could otherwise bias the $H_0$ measurement. \object{PS J0147+4630} provides 
an interesting case study in this respect.

\acknowledgments
We thank Martin Millon for making publicly available a Jupiter notebook that has greatly 
facilitated the use of the PyCS3 software. We also thank anonymous comments and suggestions 
to a preliminary version of this manuscript, which have helped us to build the current 
version. This paper is based on observations made with the Liverpool Telescope (LT) and the 
Nordic Optical Telescope (NOT). The LT is operated on the island of La Palma by Liverpool 
John Moores University in the Spanish Observatorio del Roque de los Muchachos of the 
Instituto de Astrof\'isica de Canarias with financial support from the UK Science and 
Technology Facilities Council. The NOT is operated by the Nordic Optical Telescope 
Scientific Association at the Observatorio del Roque de los Muchachos, La Palma, Spain, of 
the Instituto de Astrof\'isica de Canarias. The data presented here were in part obtained 
with ALFOSC, which is provided by the Instituto de Astrof\'isica de Andalucia (IAA) under a 
joint agreement with the University of Copenhagen and NOTSA. We thank the staff of both 
telescopes for a kind interaction. We have also used imaging data taken from the Pan-STARRS 
archive and the Barbara A. Mikulski archive for the NASA/ESA Hubble Space Telescope, and we 
are grateful to all institutions developing and funding such public databases. VNS would 
like to thank the Universidad de Cantabria (UC) and the Spanish AEI for financial support 
for a long stay at the UC in the period 2022-2023. HD acknowledges support from the Research 
Council of Norway. This research has been supported by the grant PID2020-118990GB-I00 funded 
by MCIN/AEI/10.13039/501100011033. 

%

\vspace{5mm}
\facilities{Liverpool:2m (IO:O), NOT (ALFOSC), PS1, HST (WFC3)}


\software{IRAF \citep{1986SPIE..627..733T,1993ASPC...52..173T},
 		IMFITFITS \citep{1998AJ....115.1377M},
		Python (\url{https://www.python.org/}),
                PyCS3 (\url{https://gitlab.com/cosmograil/PyCS3}),
		GRAVLENS/LENSMODEL (\url{http://www.physics.rutgers.edu/∼keeton/gravlens/})
          }



\end{document}